# HearSmoking: Smoking Detection in Driving Environment via Acoustic Sensing on Smartphones

Yadong Xie, Fan Li, *Member, IEEE*, Yue Wu,
Song Yang, *Member, IEEE*, and Yu Wang, *Fellow, IEEE*

*Abstract*—Driving safety has drawn much public attention in recent years due to the fast-growing number of cars. Smoking is one of the threats to driving safety but is often ignored by drivers. Existing works on smoking detection either work in contact manner or need additional devices. This motivates us to explore the practicability of using smartphones to detect smoking events in driving environment. In this paper, we propose a cigarette smoking detection system, named HearSmoking, which only uses acoustic sensors on smartphones to improve driving safety. After investigating typical smoking habits of drivers, including hand movement and chest fluctuation, we design an acoustic signal to be emitted by the speaker and received by the microphone. We calculate Relative Correlation Coefficient of received signals to obtain movement patterns of hands and chest. The processed data is sent into a trained Convolutional Neural Network for classification of hand movement. We also design a method to detect respiration at the same time. To improve system performance, we further analyse the periodicity of the composite smoking motion. Through extensive experiments in real driving environments, HearSmoking detects smoking events with an average total accuracy of 93.44 percent in real-time.

*Index Terms*—Smoking detection, neural networks, acoustic sensing, mobile computing

---

## 1 INTRODUCTION

WITH rapid development and great success of automotive industry, more and more vehicles have been put into use. On one hand, these modern means of transports bring people with much convenience in daily life. On the other hand, various issues related to road safety increase and arouse wide public attention. A series of efforts have been undertaken by traffic departments or government organizations to improve road safety, such as installing surveillance cameras and making traffic rules, which can in some way regulate driving behavior.

Among different undesirable driving habits, smoking is a kind of behavior that can easily distract a driver's attention and cause danger. Unfortunately, most of drivers fail to realize the risk of smoking. According to a report published by the National Institutes of Health [1], drivers who are smoking are even more distracted than people who are using cell phones, on average. The Federal Motor Carrier Safety Administration (FMCSA) also conducts its own 5-year study [2] into the dangers of smoking during driving a truck. They find that smoking is a source of distraction in 90 percent of distraction-related crashes. This equates to approximately 12,780 crashes over the

• *Yadong Xie, Fan Li, Yue Wu, and Song Yang are with the School of Computer Science and Technology, Beijing Institute of Technology, Beijing 100089, P. R. China. E-mail: {ydxie, fli, ywu, S.Yang}@bit.edu.cn.*
• *Yu Wang is with the Department of Computer and Information Sciences, Temple University, Philadelphia, PA 19122 USA. E-mail: wangyu@temple.edu.*

5-year examined period. Health damaging also should not be ignored. The British Medical Association (BMA) highlights a research indicating that because the driver smokes, the levels of toxins in a car can up to 11 times higher than that in a smoky bar. The cigarette smoke not only harms the driver himself/ herself, but also harms other passengers especially children. Many countries and areas, such as UK and Japan, have ban-smoking policy for commercial vehicles, including vans, buses, taxis and company cars [3]. Some ridesharing companies, like Uber and Lyft [4], also do not allow smoking in vehicles. But in some areas, these companies are not allowed to install monitoring equipment in cars, which makes them lack of cheap and effective detection methods. Furthermore, suppose that the detection results could be uploaded to the transportation department, then the police could further understand the driver's state when dealing with traffic accidents. Therefore, it is highly desirable to develop an easy-deployment and low-cost smoking detection system that can help companies and transportation departments to check drivers' smoking events.

There have been several existing works on cigarette smoking detection by leveraging different types of devices, such as cameras, gas sensors and Wi-Fi devices. A smoking behavior detection system [5] is proposed based on the human face analysis, which can detect whether the person in the image is smoking by locating mouth and processing white balance. Some solutions exploit the usage of various technical sensors [6], [7], such as ionization detector, photoelectric detector and gas-sensitive detector. Smokey [8], which depends on commodity Wi-Fi infrastructures, leverages the smoking patterns leaving on Wi-Fi signals to identify the smoking activity even in the non-line-of-sight and through-wall environments. However, these works suffer many problems. In particular, the methods based on computer vision heavily depend on good lighting and weather condition. Moreover, Uber and Lyft are not allowed to use the cameras and other recording devices due to privacy regulations in some areas [4]. Other methods based on specific sensors are costly or difficult to be deployed in cars.

Nowadays, smartphones become powerful with enriched inertial sensors, such as microphones, speakers and accelerators, which can be used to sense various aspects of driving conditions. Researches that focus on improving the quality of daily driving by using smartphones emerge in quantity. V-Sense [9] is a vehicle steering detection middleware that can run on commodity smartphones to detect various vehicle maneuvers, including lane-changes, turns and driving on curvy roads. Several systems put their attentions on estimating vehicle speed by using GSM signal strength traces [10], accelerometers [11] and GPS sensors [12]. However, to the best of our knowledge, a ubiquitous smoking detection system designed for driving environment is still absent. Smartphones, with their powerful capability and usability in driving, is highly ideal to act as the platform of a smoking detection system. In view of the aforementioned situations and motivations, we take the first attempt to build a novel smoking detection system, which uses acoustic sensors in smartphones, to detect drivers' smoking behaviors in real driving environment. The basic idea is that the smartphone emits acoustic signals by its speaker and receives reflected signals by its microphone, and then analyses the received signals to detect whether the driver is smoking or not. The system naturally has two advantages: smartphones are widely available and low-cost to use. In addition, leveraging acoustic sensors is a non-contact way that does not require any device to put on.

To realize this smoking detection system, we face three major challenges in practice. First, there are multiple body movements when a driver is smoking during driving, e.g., steering with one hand, holding cigarette with another hand, putting up and down the cigarette, inhaling and exhaling smoke with chest expanding and shrinking. All of these movements need to be distinguished and tracked. Second, in real driving environment, acoustic signals are easily suffered from multipath interference. Due to the limited space in the car, surfaces of various car facilities and human body can reflect acoustic signal, especially the signals of multiple reflection paths from different parts of the driver's chest when he/she is breathing. When the driver puts up and down the cigarette, the movement of the whole arm also has multipath effects. So removing different multipath interferences is necessary. Last but not least, some motions like drinking and eating have similar behavior patterns to smoking, which are very confusing to a detection system. Thus, it is a necessity to analyse the composite smoking motion to accurately detect smoking activity.

To address the above challenges, we propose a smoking detection system, named HearSmoking, which only uses acoustic sensors on smartphones in driving environments. We first analyse the smoking behaviors of 17 drivers, and find the typical smoking steps of drivers. To perceive motions, we let the smartphone speaker sends designed acoustic signals. The acoustic signals are reflected by surrounding objects and then received by the smartphone microphone. To get distances between reflectors and the smartphone, we calculate Relative Correlation Coefficient (RCC) of the collected data.

Further, we get a set of sequence profiles from RCC profiles. Each sequence profile describes distance changes between moving objects and the smartphone over a period of time. According to our observations, when a driver is smoking, his/her main moving parts are hands and chest, so HearSmoking focuses on detecting movements of hands and chest. For hand movement detection, we innovatively transform a sequence profile into a two-dimension image, and then send the image to a carefully designed Convolutional Neural Network (CNN) to identify whether there is a movement that matches the smoking hand movement pattern in the sequence profile. For chest movement detection, we perform Fast Fourier Transform (FFT) to find out waveforms in sequence profiles that fit human breath rate. Then a major breath path is selected to eliminate multipath interference. We analyse the amplitude and period of the waveform to determine whether there is a breath similar to smoking breath. If both hand movement and breath pattern fit the characteristic of those in a smoking event, we then analyse the periodicity of the detected composite motion to improve system performance. Finally, we get an analysis result whether the driver is smoking or not. To meet realistic demands, we collect training data using smartphones for 5 months to build the system model. We implement HearSmoking on different versions of Android platforms and comprehensively evaluate its performance in various environments. Experiment results show that HearSmoking is reliable and efficient in real driving environments.

Our contributions are summarized as follows:

- We study the unique patterns of smoking behaviors during driving. Based on our findings, we propose a smoking detection system, HearSmoking, which uses acoustic sensors embedded in smartphones to detect smoking events of drivers. To the best of our knowledge, we are the first to design a smoking detection system by only using smartphones.
- We divide the smoking detection into hand movement classification and respiration identification. We innovatively combine acoustic signal processing with CNN-based image classification into HearSmoking. After that, we design the methods of composite analysis and periodicity analysis to obtain the final detection result.
- We conduct extensive experiments in real driving environments. HearSmoking achieves an average total accuracy of 93.44 percent for smoking event detection.

The rest of this paper is organized as follows. We review related work in Section 2. In Section 3, we summarize the smoking steps and introduce the acoustic signals used in our work. Then we describe the system design in Section 4. Implementation and experimental results are presented in Section 5. Finally, we draw our conclusion in Section 6.

## 2  RELATED WORK

In this section, we present investigations of the existing works related to HearSmoking. To be specific, we review *driving state detection using smartphones*, *smoking detection in contact manner* and *smoking detection in non-contact manner*.

*Driving State Detection Using Smartphones.* With the increase of public awareness about road safety, many works on driving state detection using smartphones emerge to improve the quality of daily driving. SenSpeed [11] is a system for accurate vehicle speed estimation, which can estimate vehicle speed by integrating the readings of accelerometers in smartphone. D³-Guard [13] proposes a drowsy driving detection system, which leverages audio sensors in smartphones, to detect drowsy actions and alert drowsy drivers. TEXIVE [14] uses smartphones to distinguish drivers from passengers and detect texting operations during driving according to irregular and rich micro-movements of users. V-Sense [9] develops a vehicle steering detection middleware that can run on commodity smartphones to detect various vehicle maneuvers, including lane-changes, turns, and driving on curvy roads. Various kinds of works indicate the powerful capability of smartphones and embedded sensors. However, research about smoking detection in driving environment using smartphones is absent. This motivate us to propose HearSmoking to detect and alert drivers' smoking behavior.

*Smoking Detection in Contact Manner.* Technologies and studies on smoking detection using specialized devices, e.g., smart bracelets, smartwatches and chest belts, have been developed for some time. HLSDA [15] is a smoking detection algorithm, which collects various sensor data from a smartwatch and recognizes smoking behavior. PACT [16] is another wearable sensor system based on support vector machines. It detects smoking events by monitoring cigarette-to-mouth hand gestures in a contact manner. By capturing arm movements and breath puffs from 6-axis inertial sensors worn on two wrists of the user, puffMarker [17] builds a model based on 10-fold cross-validation to detect cigarette smoking. Another study [18] investigates the differences in brain signals of craving smokers, non-craving smokers, and non-smokers. This study uses data from resting-state EEG devices to train predictive models based on residual neural networks, and can distinguish the three groups. These works are all based on contact manner that need users to wear additional devices. Thus, they either cost high price to deploy or suffer inconvenience in daily using. Non-contact and device-free methods are needed for the smoking detection.

*Smoking Detection in Non-Contact Manner.* Non-contact methods are proposed by using civil cameras, gas sensors, Wi-Fi devices, etc. Smokey [8] is a smoking detection system that depends on Wi-Fi infrastructure. It leverages the smoking patterns leaving on Wi-Fi signals to identify the smoking activity even in the through-wall environments. A self-determined mechanism [19] is proposed to analyse smoking related events directly from videos by combining color reprojection techniques, Gaussian mixture models and hierarchical holographic modeling framework. Besides cameras, gas sensors are widely used. UbiLighter [20] detects cigarette smoking by using a gas sensor embedded in lighters to capture the gas from burning tobacco. A smoking monitoring method [21] uses a microphone to distinguish smoking breath from non-smoking breath. However, due to the great influence on accuracy that ambient noises would bring by only using the microphone, the system is not suitable to be used in cars. The methods based on computer vision heavily depend on lighting and weather condition. Moreover, companies are not allowed to use the cameras and other recording devices due to privacy regulations in some areas [4]. Other methods based on specific sensors are costly or difficult to be deployed in cars.

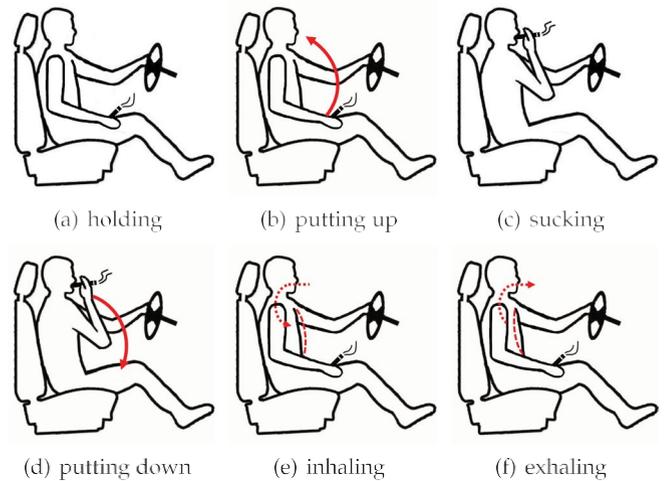

(a) holding     (b) putting up     (c) sucking

(d) putting down     (e) inhaling     (f) exhaling

Fig. 1. Typical smoking steps of drivers.

*HearSmoking.* Different from existing works, HearSmoking detects drivers' smoking behaviors only using smartphones. HearSmoking can be applied in many ways. It can help to supervise the drivers of no-smoking vehicles, such as taxis and buses. In particular, HearSmoking is very suitable to be used in Uber and Lyft, since if a passenger complaint a smoking driver, it is easier for Uber and Lyft to obtain evidence from HearSmoking. It can also work with other systems to improve driving safety. For example, it can be integrated with the ubiquitous driving modes and navigation systems on the smartphones. Furthermore, if the detection results are uploaded to the transportation department, the police can further understand the driver's state when dealing with traffic accidents.

## 3 PRELIMINARY

In this section, we present our studies and observations on smoking patterns of drivers. Then we show the design of acoustic signal according to the characteristics of smoking patterns.

### 3.1 Smoking Steps of Drivers

We first introduce the typical smoking steps of a driver during driving. After observing the smoking patterns of 17 drivers when they are driving, and referring to several on-line materials and existing works [8], we summarize the behavior patterns of smoking when a driver is driving. Generally speaking, a smoking driver usually holds the cigarette using one hand and holds the steering wheel using another hand for most of the time. Besides that, the driver puts up the cigarette to mouth and sucks the smoke intermittently. In order to facilitate detecting, we divide smoking behavior into six steps. Fig. 1 shows the detailed six steps of the smoking behavior.

- *Holding.* A smoking driver holds the cigarette for most of the time. If the driver uses the left hand to hold the cigarette, he/she may put the left hand on the car window to discard cigarette ash. If the driver uses the right hand, he/she may put the right hand on the armrest, as the ashtray is usually in front of the armrest.

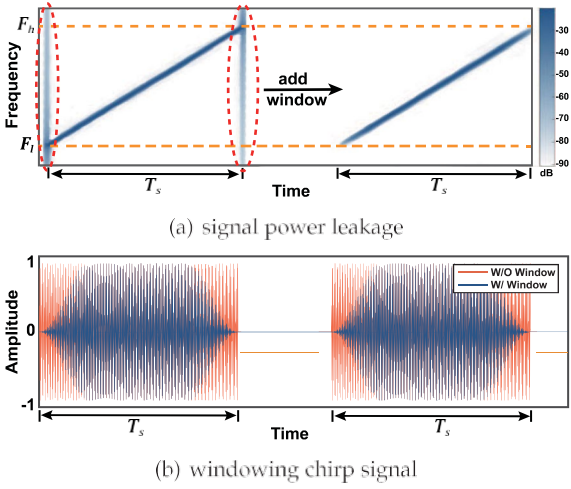

(a) signal power leakage

(b) windowing chirp signal

Fig. 2. Signal power leakage and windowing chirp.

- *Putting up*. A smoking driver puts up his/her hand and sends the cigarette to the mouth.
- *Sucking*. A smoking driver sucks little smoke into the mouth and then holds the smoke in the mouth for a moment. Note that the smoke usually is not inhaled into lungs directly.
- *Putting down*. A smoking driver puts down the cigarette and places his/her hand on the car window or the armrest.
- *Inhaling*. A smoking driver inhales the smoke into lungs with his/her hand continuously holding the cigarette.
- *Exhaling*. A smoking driver exhales the smoke with his/her hand continuously holding the cigarette.

Smoking is a rhythmic activity, in which the second step to the sixth step are carried on circularly, until the driver finishes smoking a cigarette. In HearSmoking, a round from the second step to the sixth step is regarded as a smoking motion. When a driver smokes during driving, his/her hands and chest all move regularly with different distance ranges. Therefore, to detect smoking events in driving environments, we turn the smartphone into an active sonar by using acoustic sensors, i.e., the speaker and the microphone embedded in the smartphone. The acoustic sensors are highly applicable to track the moving objects like hands and chest at different distances.

## 3.2 Acoustic Signal Design

The basic idea of movement detection is to estimate distance $d$ between an object and the smartphone using acoustic signals. The smartphone transmits and receives a signal and then computes the time delay $t$ to estimate the distance by $d = t \cdot v$, where $v$ denotes the speed of acoustic signal. To build the active sonar on the smartphone, we first need to design matching acoustic signals. The commonly used acoustic signals include signal with fixed frequency [22] and Frequency Modulated Continuous Wave (FMCW) [23]. Compared with the fixed frequency signals, FMCW can facilitate the RCC calculation and also tolerate the frequency selective fading of audio signals due to multipath effect in driving environment. So we choose FMCW, of which the transmitted frequency increases linearly cross time, as shown in Fig. 2a. In the time

domain, the chirp signal of FMCW is formed as

$$s(t) = A\cos\left[2\pi\left(f_c t + \frac{B}{2T_s}t^2 - \frac{NT_s}{2T_s}\right)\right];\quad(1)$$

where $t \in [NT_s - \frac{T_s}{2}, NT_s + \frac{T_s}{2}], N \in \mathbb{Z}$. In a chirp signal, $A$ denotes the amplitude of the signal, and $f_c = \frac{F_h + F_l}{2}$ denotes the carrier frequency. $F_l$ and $F_h$ represent the lowest and highest limit of frequency respectively. $B = F_h - F_l$ is the bandwidth, and $T_s$ is the sweep time.

We take several factors into consideration in the signal design. First, the FMCW should be as inaudible as possible to human ears for avoiding disturbing the driver in normal driving. Second, in order to reduce interferences that echoes from objects overlap with emitting signal in the time domain, the time duration of the FMCW should be short enough. The upper limit frequency of human hearing is about 20 kHz [24]. To make the FMCW inaudible to users, the lowest limit of frequency $F_l$ has to be higher than 20 kHz. According to Nyquist sampling theorem, and considering that the highest sampling rate supported by most of mainstream smartphones is 48 kHz, the highest limit of frequency $F_h$ has to be lower than 24 kHz. Therefore, comprehensively considering the above two limitations, we set $F_l = 20$ kHz and $F_h = 22$ kHz. Each chirp signal in FMCW contains 64 samples, corresponding to $T_s = 1:33$ ms under the typical 48 kHz sampling rate of the smartphone speaker. Furthermore, the separation between two continual chirp signal is set to 400 samples, corresponding to $T_i = 8:33$ ms. This length of separation ensures that all echoes within the range of 1:4m can be captured without being inter-fered by the next emitting chirp signal. Compared with con-tinuous transmitting acoustic signal without separation, the chirp signal with separation can also reduce the computa-tional burden and energy consumption of the smartphone.

In the usage of FMCW signal, there exists a problem caused by imperfection of commercial speakers, which is the power leakage between two continuous chirp signals. The left part of Fig. 2a shows the spectrogram of the emit-ting signals. Strong power leakage is notable at the begin-ning and the end of chirp signals due to frequency hopping, making FMCW signals audible to human ears. To reduce this influence, we add a tapered cosine window on the emit-

$$w(x) = \begin{cases} \frac{1}{2}\left[1 + \cos\left(\pi\left(\frac{2x}{uL} - 1\right)\right)\right]; & \text{if } 0 \le x < \frac{uL}{2}; \\ 1; & \text{if } \frac{uL}{2} \le x \le L - \frac{uL}{2}; \\ \frac{1}{2}\left[1 + \cos\left(\pi\left(\frac{2x}{uL} - \frac{2}{u} + 1\right)\right)\right]; & \text{if } L - \frac{uL}{2} < x \le L: \end{cases}\quad(2)$$

Particularly, the amplitude value of $x$th sample in a chirp sig-nal multiplies window function $w(x)$ to reduce power leak-age, where $L$ is the length of a chirp signal (here is 64) and $u$ is the ratio of cosine-tapered section length to the entire window length (here is 0.5). The signals before and after adding tapered cosine window are shown in the Figs. 2a and 2b, in terms of frequency and amplitude respectively.

## 4 SYSTEM DESIGN

In this section, we present the design of HearSmoking in details. We first show the overview of HearSmoking, includ-ing online and offline processes. Then we present how to detect hand movements and respiration by leveraging acoustic signals.

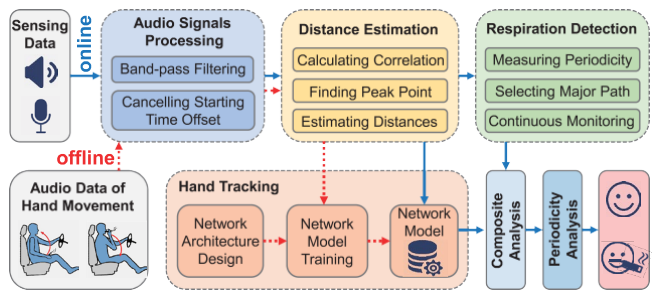

Fig. 3. System overview.

### 4.1 System Overview

Fig. 3 shows the overview of HearSmoking. The whole system can be divided into offline part and online part. We use solid lines with arrowhead to denote the online process and dotted lines with arrowhead to denote the offline process.

#### 4.1.1 Offline Process

In the offline phase, HearSmoking first emits the designed FMCW signals and receives the reflected signals during driving. Then HearSmoking preprocesses the received signals by a band-pass filter, which can remove out-of-band interference and the human voice to meet the privacy requirements. We design a window-based method to cancel starting time offset. After that, HearSmoking calculates RCC of received signals and emitting signals to find static and moving objects by ana- lysing peaks in RCC profiles. Then we can obtain the distance changes over time of these static and moving objects, denoted as distance change sequence. To train model for hand move- ment detection, we manually distinguish smoking hand movement according to their features and recorded videos, then the distance change sequences are divided into sequence profiles. We label each sequence profile with indication whether it contains smoking hand movement. The labeled sequence profiles are sent into a carefully designed CNN for model training. Finally, HearSmoking gets a network model for

#### 4.1.2 Online Process

In the online phase, HearSmoking continuously emits and receives the designed FMCW signals during driving. The received signals are filtered and sequence profiles of distance change are obtained by the same method in the offline phase. The obtained sequence profiles are then sent into the trained network to detect whether there are smoking hand movements. At the same time, the distance change sequences are analysed for respiration detection. Specifically, at the beginning when the smartphone is put in the car, HearSmoking first finds sequences of chest movement related to breath according to their natural periodicity. Due to the multipath effect, we choose the sequence with the most obvious respiratory characteristics as the major path signal. Then HearSmoking only detects breath signals on that major path. Once the network detects hand movements that may relate to smoking, the short period of breath signal after hand movement is extracted for a composite analysis. In order to improve detection accuracy, if the features of hand movement and breath fit the characteristic of a smoking motion, then HearS- moking conducts a periodicity analysis to check whether the motion fits the periodic characteristic of a smoking event. Finally, HearSmoking can detect whether the driver is smok- ing or not when driving.

### 4.2 Movement Distance Estimation

The basic idea of smoking detection is that the smartphone emits high-frequency acoustic signals and receives reflected acoustic signals to capture different movements of the driver. How to estimate the distance changes of hand movements and chest movements from collected signals is one of key challenges in HearSmoking.

#### 4.2.1 Signal Preprocessing

After collecting signals, we first remove noises at undesired frequencies by using a band-pass filter and remain signals with the frequency range from 20 kHz to 22 kHz. The filter can also remove the human voice to meet the privacy requirements. Since FMCW signals are transmitted periodically from the smartphone speaker, it is necessary to determine where the first reflected FMCW signal is in the received signals collected by the microphone. Ideally, if the speaker and the microphone start to work simultaneously, the propagation time from smartphone to a reflector can be estimated directly. However, it is hard to perfectly synchronize the start time of the speaker and the microphone due to hardware limitation. As the distances between the speaker and the microphone of most smartphones are less than 15 cm, the propagation time of Line-of-Sight (LOS) signal can be ignored. So, we set the time of receiving the LOS signal as the starting point. Considering that there is a short-time offset between the emitting signal and its corresponding received LOS signal, we apply a window with length of 64 samples to find this start-time offset. Using $r$ to denote the emitting signal and $r^o(n)$ to denote the received signal in the $n$th window, HearSmoking estimates subsample delay $d(n)$ with phase slope changing [25] in the frequency domain

$$d(n) = \min_d k \, \|F(r) - F(r^o(n)) + 2\pi fd\, k;\| \qquad (3)$$

where $F$ is Fourier transform, and $f$ is a parameter related to the window length. Commonly used method to find start-time offset is to use a sliding window that slides one sample at each time. Similarity between $r$ and $r^o(n)$ is calculated at each sliding. Finally the maximum similarity is selected and the corresponding $n$th sample is the starting point of FMCW signal in the received signals. However, this method leads to large computing load, so we apply another light-weighted method to reducing computing load. The main idea of our method is when the window is at the initial position, we calculate the value of $d(n)$ and then the window shifts $d(n)$ samples. We repeat the above steps until $d(n) < 1$ or $d(n) > 64$, and the stopped position of the window is the starting point of FMCW signal in received signals. Note that this process only needs to be calculated once at the beginning of the driving, and each time after the window is moved. This process is shown in Fig. 4a. We can see from Fig. 4b that the commonly used method needs to calculate 650 times to find the starting point. But our modified method only needs 20-time calculation, as shown in Fig. 4c.

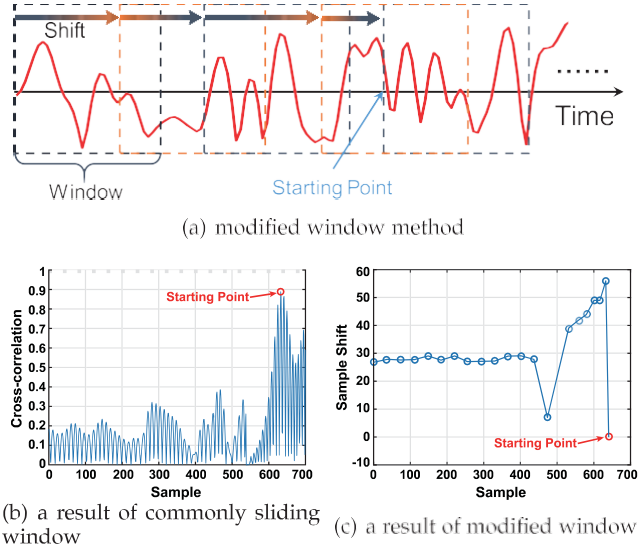

(a) modified window method

(b) a result of commonly sliding window

(c) a result of modified window

Fig. 4. Start-time offset cancellation.

#### 4.2.2 Distance Estimation

Once we find the starting point, we can calculate the distances of the obstacles that reflect the signal. The received audio signals contain many signals reflected by various obstacles around the smartphone, which are considered as multipath signals. To find signals reflected by our targets, i.e., hands and chest of the driver, HearSmoking needs to distinguish obstacles' movements at different distances. An effective way is to compare the correlation between the emitting signal and the received signals from the starting point. In specific, a sliding window with length of 64 samples is added on the received signals, and then the correlation is estimated by computing RCC between the emitting signal and the received signals in the sliding window. RCC is defined as

$$R(r, r^0) = \frac{Cov(r, r^0)}{s_r s_{r^0}} \left( \frac{r^0}{r} \right),$$  (4)

where $r$ and $r^0$ denote the emitting signal and the received signals in the sliding window, respectively. $Cov(r, r^0)$ is the covariance of $r$ and $r^0$, while $s_r$ and $s_{r^0}$ respectively represent the standard deviations of $r$ and $r^0$. Moreover, $r$ and $r^0$ are the amplitude means of $r$ and $r^0$ in length of 64 samples. Fig. 5 shows RCC profiles from time slot $t_1$ to $t_5$ in real detection environment. Each peak in the RCC profile represents an obstacle which reflects audio signals. If two peaks overlap at two consecutive times, it means the corresponding obstacle is static relatively to the smartphone. Otherwise, it is a moving object such as a driver's hand. As we can see from the Fig. 5, when a driver's hand moves from $0.7$m at $8.2$s to $0.735$ m at $9.3$ s, there are 5 corresponding moving peaks in RCC profiles.

In practice, HearSmoking constantly collects acoustic signals and generates RCC profiles of each moment. Since the time interval between two moments is very short, the moving distance of an obstacle between these two moments is also very short. In a nutshell, we regard the two closest peaks in two adjacent RCC profiles as the moving distance of the same obstacle. Then distance changes of every obstacle over time are extracted from these RCC profiles. We select a sequence profile of distance change that are collected in real detection environment, as shown in Fig. 6. We focus on obstacles at the distance less than 1m due to the limited driving space of vehicles. Those obstacles related to sequences with nearly no fluctuation are static objects ($S_6$ in Fig. 6), while other obstacles with fluctuating sequences are moving objects. So, sequences that are corresponds to static objects can be removed by calculating the movement range. It should be noted that there are also some sequences with few durations, this is because the driver or co-pilot makes big movements, or the car sometimes violently vibrates when passing speed bumps. Thus we remove sequences with durations less than 3 s. Among the sequences of moving objects, HearSmoking can distinguish sequences related to chest movements and hand movements according to their unique movement patterns. The sequences of chest movements and hand movements are used to track hands and detect respiration.

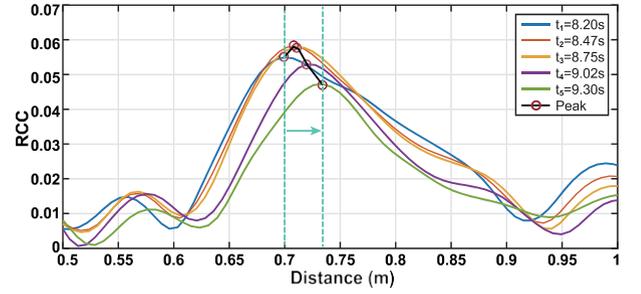

Fig. 5. RCC profiles between $8.2$s and $9.3$s.

### 4.3 Respiration Detection

To detect the driver's respiration, we need to determine which sequence is related to chest movement. When a driver gets into the car, he/she usually first drives off from the parking lot. This process may contain drastic body motions like sharply turning the steering wheel or turning around head, which could cause great disturbance on respiration signals. Thus, we select a one-minute received signal $\mathbf{F}_0$, which starts at the time moment that is two minutes after the starting point, to analyse respiration signal. The typical chest $1 cm$ during normal breath [26]. However, our study find that the depth of smoking breath is usually less than $2.5$ cm. So, after getting the sequence profile (with useless sequences removed) of $\mathbf{F}_0$, HearSmoking selects sequences with amplitude change range less than $2.5$ cm. Then Fast Fourier Transform is applied to each selected sequence to find the components whose frequency is mainly within normal respiration range ($0.16$ Hz-$0.6$ Hz) [27]. Fig. 7 shows the result of selected sequences after FFT. We can see that the main components of $S_1$, $S_2$ and $S_3$ are obvious at around $0.35$ Hz, while $S_7$ has no obvious main component. Due to multipath effect, there are three sequences (i.e., $S_1$, $S_2$ and $S_3$) that related to breath. For better performance, we select the sequence with the largest FFT amplitude ($S_2$) as major breath path, which indicates more obvious feature for detection. If HearSmoking does not find major breath path in $\mathbf{F}_0$, it will continue to detect breath until it finds major breath path. Once HearSmoking determines the major breath path, it analyses respiration on major breath path. Note that the driver's main body usually stays static during stable driving, thus the starting point of received signals and major path of respiration signals only need to be determined once at the beginning of detection, and each time the smartphone is moved.

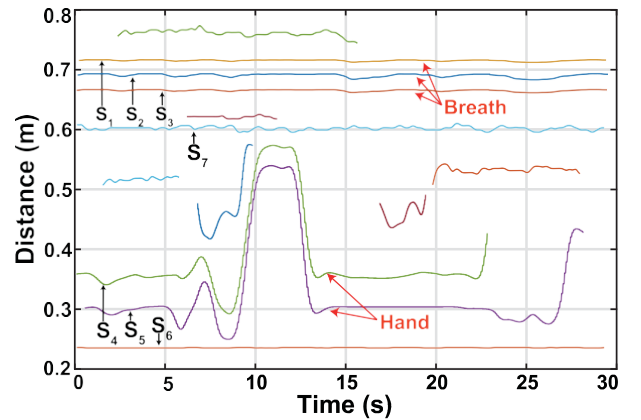

Fig. 6. A sequence profile of distance change.

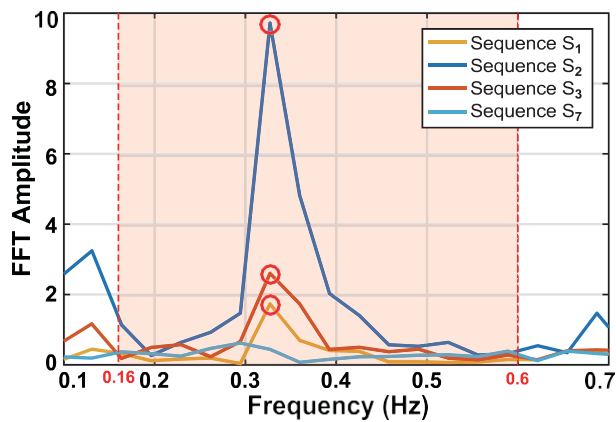

Fig. 7. FFT result of selected sequences.

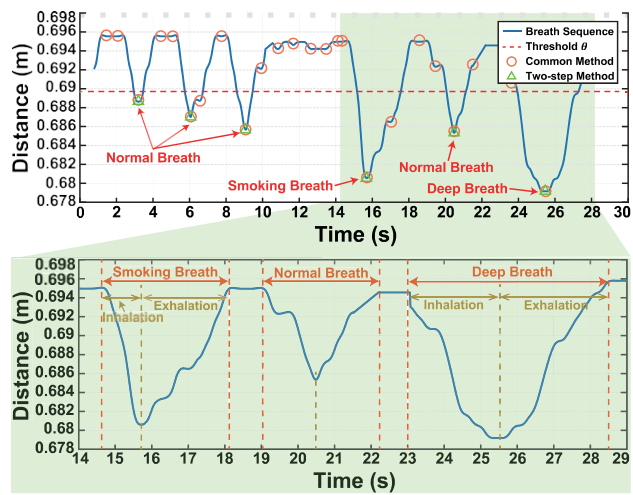

Fig. 8. Understanding trough selection.

In order to distinguish smoking breath from non-smoking breath, it is necessary to obtain the depth and duration of each breath. The smartphone is placed in front of the driver most of the time, so when the driver inhales, the distance between the chest and the smartphone is shortened, and when the driver exhales, the distance between the chest and the smartphone is enlarged. The breath sequence can be approximated as a periodic sinusoidal wave, in which the trough amplitude represents breath depth. Common trough detection method identifies the transition point at which the signal changes from a decreasing to an increasing trend. However, such method would lead to plenty of useless troughs in our respiration detection. Thus, we design a two-step trough selection method to compute the amplitude and periodicity of the breath sequence.

- *Step 1:* Set a threshold on the distances of the detected troughs. Specifically, we run a common trough detection method on a one-minute breath sequence to obtain an initial set of troughs. Then the mean and standard deviation of the trough distances, denoted as $m_p$ and $s_p$, are calculated. After experiments, we set threshold as $u = m_p - \frac{1}{4} s_p$ and remove those troughs whose distances are larger than the threshold.

- *Step 2:* Compare trough distances. Considering that a single breath usually lasts 2s to 6s [27], we set the first trough as current trough and check whether the time between it and the second trough is less than 2 s. If so, then we compare the distances of the first trough and the second trough and remove the trough with larger distance, and further set the remaining trough as current trough. Otherwise, we reserve both the first trough and the second trough, and further set the second trough as current trough. This step is repeated until all the troughs are

Through running our trough selection method, we can get lots of information about the depth and duration of each breath. Fig. 8 shows a part of breathing sequence in the major path ($S_2$ in Fig. 6), including normal breath, deep breath and smoking breath. We also show the results of common trough detection and our two-step trough selection.

We compare the depth and duration of smoking breath with those of normal breath and deep breath by conducting an extra experiment. We analyse three types of breath in Fig. 8. The depth of normal breath is about $1 cm$ [26], while the depth of deep breath or smoking breath is generally larger than $1.4$ cm. It is obvious that smoking breath and deep breath have larger amplitudes and longer durations than those of normal breath. The biggest difference between smoking breath and deep breath is that the inhalation duration is as long as the exhalation duration in a deep breath, while the inhalation duration is shorter than the exhalation duration in a smoking breath [8]. Based on the above findings, HearSmoking distinguishes smoking breath from non-smoking breath by calculating their inhalation durations, exhalation durations and depths. We design a simple method to estimate the durations since they cannot be obtained directly. Specifically, we select the top ten highest peaks in the major path signal $\mathbf{F}_0$ and calculate the mean distance $d_0$ of the ten peaks. Then depth of each breath is obtained by calculating the difference between $d_0$ and its trough distance. If the depth of a breath exceeds $1.4$ cm, we consider this breath could be a smoking breath. To reduce the interference of deep breath, we also check whether the breath is generally symmetric. We calculate two absolute average slopes of the one-second signal before and after the trough point. If the difference between the two slopes is greater than a threshold, the breath is regarded as a smoking breath.

TABLE 1
Information of Volunteers

| Gender | 10 male, 3 female |
|---|---|
| Age | 23 - 55 |
| Stature | 157 - 183 cm |
| Weight | 41 - 77 kg |
| Driving years | 1 - 27 |
| Smoking years | 1 - 21 |
| Smoking handedness | 4 left, 9 right |
| Vehicle types | 3 electric, 10 gasoline |

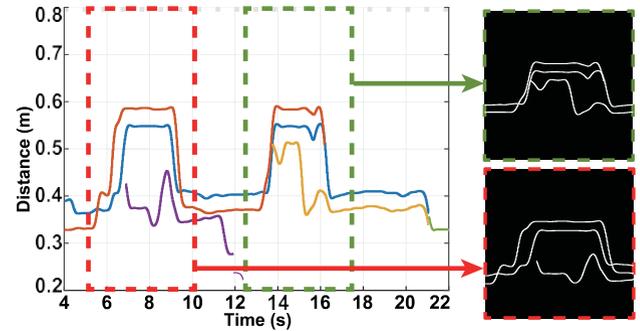

Fig. 9. Feature matrix transformation.

### 4.4 Hand Movement Detection

Drivers have different preferences of placing their hands when driving, such as using one hand to operate the steering wheel and putting another hand on the car window or the armrest. Besides, a driver also has different moving patterns when he/she uses different hands to smoke. Thus, it is hard to distinguish various kinds of smoking hand movements from other hand movements by observing frequencies and amplitudes of reflected signals. There are many popular methods for movement classification, such as K-Nearest Neighbours (KNN), Support Vector Machine (SVM) and Convolutional Neural Network. From previous research in human activity recognition [28], [29], the performance of CNN is shown to be robust against different movement speeds and amplitudes. Similarly, in our paper, the speed and amplitude of the same movement performed by different drivers are also different. So we address this problem by applying a CNN to train a classifier in the offline phase, and then the classifier is used for hand movement detection in the online phase.

To collect training data, we develop an Android application to generate and collect acoustic signals. Taking both generality and diversity of smoking behaviors into consideration, we recruit 13 drivers who have more than 1-year smoking experience to take part in our experiment. The information of the 13 drivers is shown in Table 1. The drivers smoke in daily drivings as they used to do, with their own smartphones collecting acoustic data. We also equip each vehicle with a camera to capture the driving activities of the drivers. The experimental scenarios include urban road (at peak time and off-peak time), highway and country road. After 5-month collection, we obtain 9,438 smoking motions in real driving environment. Besides real data, we also invite extra 6 volunteers to collect 3,672 simulated data as a supplement. To be specific, the volunteers imitate smoking motions with different kinds of postures when sitting in a vehicle. Finally, we collect about 13,000 smoking motion data to train the CNN. All procedures are approved by the Institutional Review Board at Beijing Institute of Technology. After that, we select and cut the smoking data into isometric segments with length of 5 s. Each selected segment contains the entire process of putting up and down hand in a smoking motion. Then, these data segments are transformed into sequence profiles by using the method mentioned in Section 4.2. Respiration sequences and too short sequences are removed in each sequence profile according to the method mentioned in Section 4.3. We label each sequence profile with the indication that whether it contains smoking hand movements or not, and then the labeled sequence profiles are used for training model.

CNN is one of the most popular algorithms in the field of deep learning. It is particularly applicable for image classification [30], which can achieve a good balance between accuracy and computing load. Considering that the sequence profile is a two dimensional feature that can be viewed as an image, we transform each sequence profile into a 96 × 96 bicolor image (with sequence line in white and other space in black for convenience of calculations), defined as a feature matrix. The transformation is shown in Fig. 9. We design a 4-layer CNN to train the feature matrices, as CNN can reduce the influence of different initial positions of hands, and has good anti-interference ability to multipath effect. Fig. 10 shows the architecture of CNN. It mainly consists of two convolution layers with maxpooling, one fully connected layer and one softmax layer. The sizes of the two convolutional and two maxpooling kernels are 5 × 5, 3 × 3, 2 × 2 and 2 × 2, respectively. The convolution layers are the most important among all layers. They extract deep-level features by convolving the input matrix with different filters. The weights of the filters are initialized by a Gaussian distribution. Then they can automatically update during back propagation. Convolution operation is the basic operation of image processing. Formally,

$$y_{mn} = s\left( \sum_{j=0}^{J} \sum_{i=0}^{I} c_{m+i, n+j} W_{ij} + b \right); m \geq 0, M; n \geq 0, N;$$

(5)

where $c$ is the input matrix, $W$ is the $J \times I$ kernel, and $b$ is the bias. $s(\cdot)$ is the activation function and $y$ is the output. We select Rectified Linear Unit (ReLU) as the activation function to increase the nonlinear properties. A maxpooling layer is added after each convolution layer to reduce feature size. Then, we add a flatten layer and a fully connected layer with 128 units. Finally, the result is produced through a softmax layer. We use $P^K$ to denote the probability of $K$th type result, then a class probability vector $P ¼ fP^1; P^2; \ldots ; P^K g$ is calculated as

$$P ¼ sð W^T h þ bÞ; \qquad (6)$$

where $sð Þ$ is a softmax function, $W^T$ is a weight matrix, $b$ is a bias vector and $h$ is the output of the previous layer.

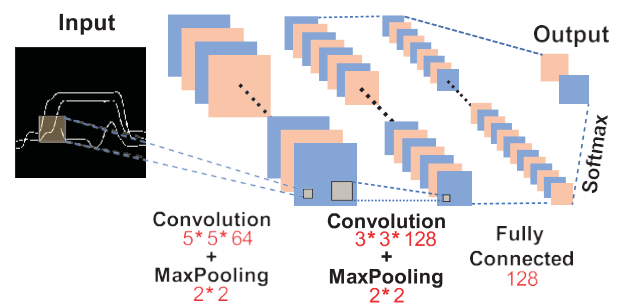

Fig. 10. Architecture of CNN.

In the online phase that trained CNN detects hand movements of smoking, HearSmoking cuts the collected driving data into isometric segments with length of $5s$. Every continuous segments have an overlap of $2s$ to guarantee that most of the data that related to putting up and putting down hands in one smoking motion can be divide into one segment. Then, these data segments are transformed into sequence profiles using the same method that in the offline phase, and the sequence profiles are then transformed into $96 \times 96$ feature matrices. These feature matrices are sent to the trained CNN for classification. We can get the result whether each segment contains data of smoking hand movements. Once CNN detects there is a segment containing data of smoking hand movements, HearSmoking can determine the approximate time when this smoking hand movement starts and ends. The classification result and the time are then used in composite analysis and periodicity analysis.

### 4.5 Periodicity Analysis for Composite Smoking Motion

As smoking is a rhythmic activity with composite motions, only detecting hand movement or respiration would lead to mistakenly recognizing other activities as smoking activity, such as eating and drinking during driving. Thus, we analyse the whole process of a smoking activity to reduce interference of similar activities. According to our description in Section 3, the inhalation and exhalation are closely behind putting up and then putting down hands. Based on this phenomenon, HearSmoking first conducts a composite analysis. Specifically, in the online phase of detection, once the network identifies a set of smoking hand movements, consisting of putting up and then putting down hands, a period of signal that is closely behind the hand movement is selected for respiration detection. If HearSmoking detects there is an inhalation and then an exhalation of smoking breath in the selected signals, it can be viewed that the driver has finished a smoking motion once. As a monitoring system, false alarm is desired to be as less as possible. Thus, to reduce false alarm, HearSmoking next conducts a periodicity analysis to detect whether there are another $1$ times smoking motions after the first detected smoking motion. Through experiment, we set to 3 for better system performance. It should be noted that the time intervals between any two continuous smoking motions usually have similar length. So until HearSmoking detects three periods of smoking motions and the two time intervals have similar length, it determines that there is a smoking event and sends out an alert. Otherwise, HearSmoking takes the detected smoking motions with different intervals as invalid motions.

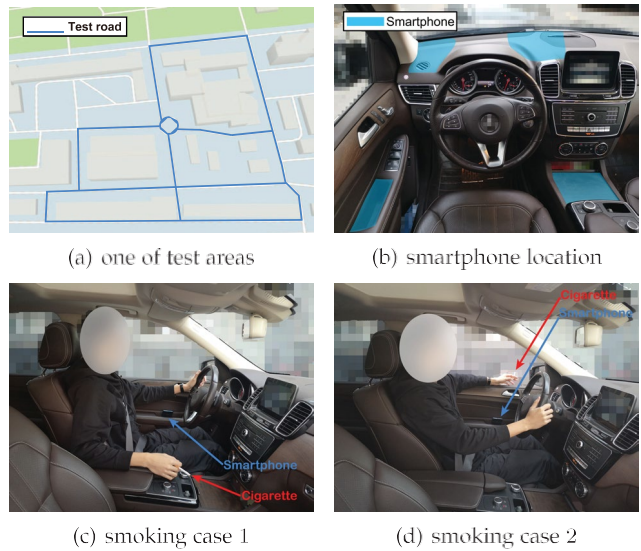

(a) one of test areas      (b) smartphone location

(c) smoking case 1      (d) smoking case 2

Fig. 11. Test scene.

## 5 IMPLEMENTATION AND EVALUATION

In this section, we introduce the implementation details and provide the evaluation results.

### 5.1 Evaluation Methodology
#### 5.1.1 Evaluation Setup

To implement HearSmoking, we develop a prototype and install it on different Android smartphones, including SAMSUNG GALAXY Note3, HUAWEI P20, HTC One M9 and HUAWEI Mate 20 Pro. The 13,000 smoking motions we collect in Section 4.4 are only used for training CNN. For evaluation, we invite another 6 drivers to collect test data, including 4 males and 2 females, for over 2 months. They vary in age (23-52 years old), stature (160-186 cm), weight (42-86 kg) and BMI (16.4-25.3). They drive different vehicles, including electric vehicles and gasoline vehicles. During testing, the drivers smoke in their own way when driving. In order to get ground truth, each vehicle is equipped with a camera to capture drivers' actions. Finally, we collect about 3,200 smoking motions as the smoking test data. In addition, non-smoking test data include several motions that drivers often appear when driving, such as operating the steering wheel, changing gear, eating and drinking. In addition, the non-smoking test data also include the data of drivers driving smoothly, which does not include the motions. Fig. 11 shows our test scene. One of the urban test areas is shown in Fig. 11a, where the drivers can choose any routes as they like. We observe that most of the drivers prefer to put their smartphones in 4 locations in vehicles, which are marked in Fig. 11b. Figs. 11c and 11d show two cases where the driver smokes with his/her right hand and left hand. All procedures are approved by the Insti- tutional Review Board at Beijing

#### 5.1.2 Evaluation Criterion

To evaluate the performance of HearSmoking, we define metrics from two aspects. In order to explain these two metrics more intuitively, we assume that a driver actually smokes for $A$ times, and HearSmoking detects $D$ times of smoking events, $C$ times of which are correctly detected.

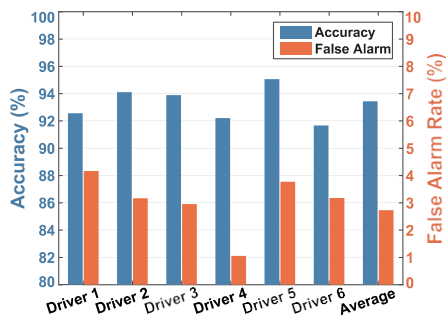

Fig. 12. Overall performance of 6 drivers.

- Accuracy: The percentage of cases where HearSmoking correctly detects the smoking events in all actual smoking events, i.e., $Accuracy = \frac{C}{A}$.
- False Alarm Rate: The percentage of cases where HearSmoking make false alarms in all the detected smoking events, i.e., $FalseAlarmRate = 1 - \frac{C}{D}$.

## 5.2 System Performance Evaluation

### 5.2.1 Overall Performance
We first evaluate overall performance of HearSmoking in real driving environment. Fig. 12 shows the detection accuracy for the 6 testers. The smoking patterns and driving habits of the testers are different from each other, leading to different smoking motion. However, it can be seen that HearSmoking achieves an average accuracy of 93.44 percent and no less than 91.67 percent accuracy for each tester. Besides, the average false alarm rate is 2.79 percent, and the highest false alarm rate among those of 6 testers is 4.11 percent, which is accept- able for a smoking detection system. The evaluation results indicate that HearSmoking is effective in detecting smoking activities in real driving environment.

### 5.2.2 Impact of Composite Analysis and Periodicity Analysis
To verify the necessities of the composite analysis and the periodicity analysis, we compare our method with three methods that are denoted as *Hp*3, *Bp*3 and *HBp*. *Hp*3 means only detecting three periods of smoking hand movements and *Bp*3 means only detecting three periods of smoking breath. *HBp* means only detecting one period of hand movements and one period of breath. Similarly, our proposed method can be called as *HBp*3. Fig. 13 shows comparison results. *Hp*3 can achieve an average accuracy of 95.07 percent but a high false alarm rate of 7.46 percent. *Bp*3 has a better false alarm rate of 5.91 percent, but it is still a little high. *HBp* performs slightly better than *Bp*3 and achieves average false alarm rate of 4.35 percent. It is obvious that *HBp*3 outperforms the three comparison methods. *HBp*3 achieves much lower false alarm rate while only sacrificing a little detection accuracy. Therefore, it is quite necessary to apply composite analysis and periodicity analysis in smoking detection.

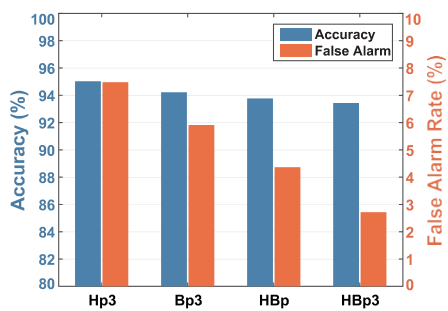

Fig. 13. Impact of composite analysis and periodicity analysis.

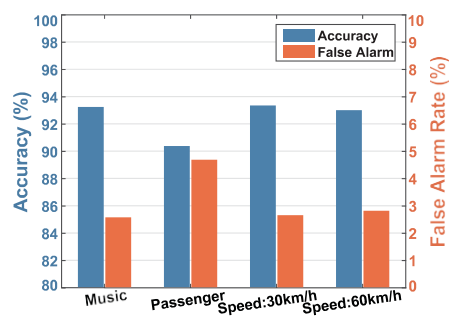

Fig. 14. Impact of driving environment.

### 5.2.3 Impact of Driving Environment
We also conduct our evaluation experiments in different driving environments, such as driving while listening to music, driving with a co-pilot and driving on different speeds. As Fig. 14 shows, listening to music has nearly no impact on the system accuracy, which is as high as 93.24 percent, as HearSmoking leverages high-frequency acoustic signals that are very easily to be separated from ambient noises. HearSmoking performs slightly worse when there is a passenger sitting on the co-pilot seat. This is because the movements of the co-pilot would be captured by HearSmoking and mistakenly identified as a smoking motion. We find that most of the time, the distance between the smartphone and the driver is smaller than that between the smartphone and the co-pilot. Therefore, if the system detects two different breathing waveforms, we default that the breath closer to the smartphone belongs to the driver. This simple measure can reduce most of the influences of the co-pilot. HearSmoking in situations of driving on 30 km=h and 60 km=h has similar accuracy and false alarm rate, which indicates that normal driving speeds have little impact on HearSmoking. In general, HearSmoking works reliably in different driving environments.

### 5.2.4 Impact of Relative Location and Angle
The relative location of the driver and the smartphone is an important impact factor in smoking detection. Different drivers like to put their smartphones on different specific locations of the car. These locations include dashboard (left), dashboard (right), door panel and storage space near gearstick. Using right hand or left hand to hold cigarette also cause differences in detection. Thus, we study the system performance in situations where the smartphone places on four typical locations and the driver holds cigarette by different hands. Fig. 15a shows the system accuracy under different smartphone locations. It can be seen that the accuracy is slightly higher when the smartphone is placed on higher locations. Specifically, putting the smartphone on the dashboard achieves the best system performance, as the smartphone directly faces the driver, and there are no obstacles between hands and chest of

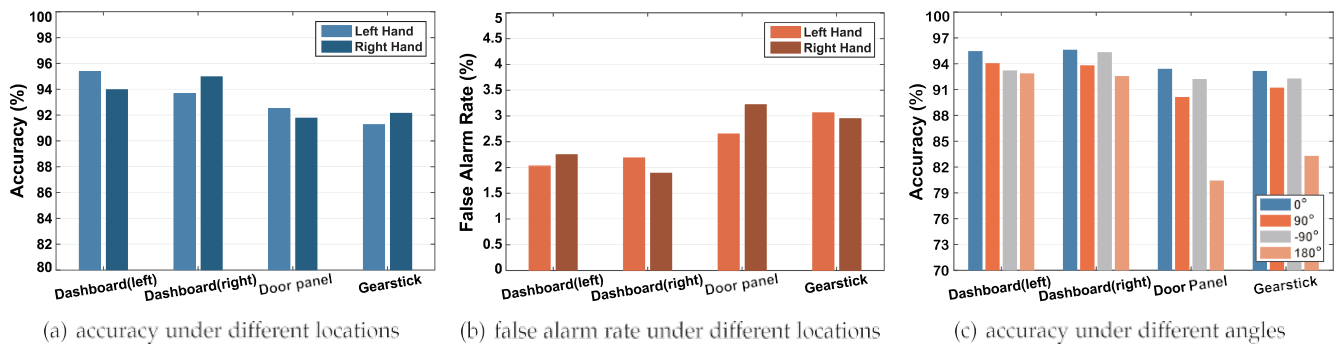

Fig. 15. Impact of relative location and angle.

(a) accuracy under different locations    (b) false alarm rate under different locations    (c) accuracy under different angles

the driver and the smartphone. Putting the smartphone on the storage space leads to the lowest accuracy, since the smartphone is relatively far away from the driver's hands and more easily influenced by the co-pilot and driver's leg movements. But no matter where the smartphone is, the accuracy are no less than 91.19 percent. Fig. 15b shows the false alarm rate under different smartphone locations, which has the similar results as those in Fig. 15a. The false alarm rate are no more than 3.23 percent.

Besides location, the placement angle of the smartphone may also have an impact on system performance. So we conduct experiments under different angles. We find that most speakers and microphones are on the bottom of smartphones, so we define the angle as 0 when the bottom of the smartphone is facing the driver. When the smartphone turns left, the angle decreases, otherwise the angle increases. We measure the accuracy of 4 angles at 4 locations, and the results are shown in Fig. 15c. We can see that when the smartphone is on the dashboard, our system has good accuracy at each angle. But when the smartphone is on the door panel or storage space near geartick, our system has relatively low accuracy at 180 (i.e., bottom of the smartphone faces away from the driver). Because speakers and microphones of smartphones are often completely blocked by surrounding objects. However, we observe that this way of placing smartphones rarely occurs, since it is not convenient for drivers to check smartphones. In general, our system works reliably in most angles.

### 5.2.5 Impact of Clothing

We evaluate how clothing of the driver influence the performance of HearSmoking. The experiment is conducted under four typical clothing: shirt, sweater, shirt þ sweater, shirt þ sweater þ coat. The result is shown in Fig. 16a. We can see that the less clothes the driver wears, the higher accuracy HearSmoking achieves. This is because clothes, especially loose clothes like coats, could block acoustic signals and partly hide the body movement brought by breathing. However, even the lowest accuracy brought by 'shirt þ sweater þ coat' is no less than 91.08 percent, which is high enough for detection. Moreover, we can see that clothing has little impact on false alarm rate. The reason may be that loose clothes just hide useful signals, especially breath signal, but not produce confusing signals. As mentioned in Section 4.3, HearSmoking determines the major path of breath signal through analysing a one-minute acoustic signal. However, due to the influence of driver's clothing, it may need to analyse another one-minute signal again to find the major path. Thus, we also evaluate how many times HearSmoking needs to analyse one-minute acoustic signal to find the major path. Fig. 16b shows the CDF of analysis time. We can see that more than 90 percent major paths can be found within 3 times of analysis even under the thickest clothes.

### 5.2.6 Impact of Road Type

The driver may have various behaviors when driving on different types of roads with different road conditions, which could also influence the performance of HearSmoking. We conduct experiments on urban road (at peak time and off-peak time), highway and country road. Fig. 17 shows the system performance under different road types. It can be seen that when the driver is driving on the urban roads during off-peak time and highways, HearSmoking has higher accuracy and lower false alarm rate. This is because when driving on these two kinds of roads, the driver rarely has other motions, and the road conditions are better, which leads to a better system performance.

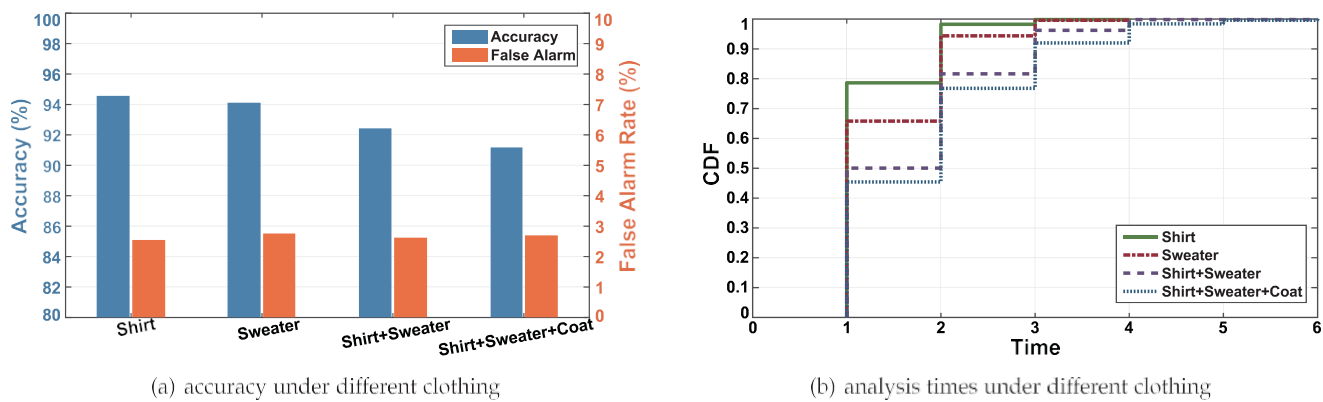

(a) accuracy under different clothing    (b) analysis times under different clothing

Fig. 16. Impact of clothing.

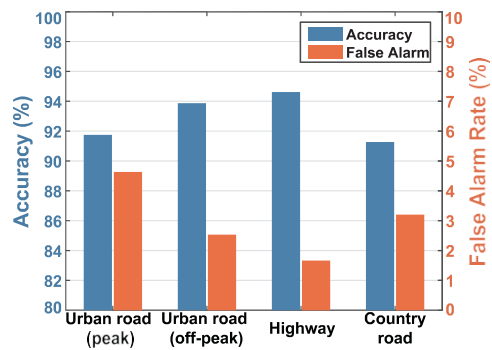

Fig. 17. Impact of road type.

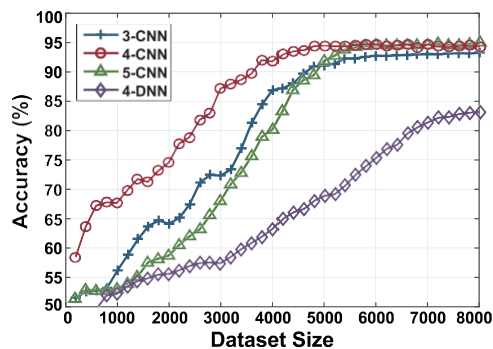

Fig. 19. Impact of network architecture and dataset size.

However, on urban roads during peak time, the driver frequently has motions like brake or turn, while road conditions of country roads are usually poor. So driving on these two kinds of roads would cause a little lower system accuracy. On the whole, the accuracy are no less than 91.28 percent, the false alarm rate are no more than 4.68 percent, showing a good robustness in various road types.

### 5.2.7 Impact of Period Detection Times

In order to select the best for HearSmoking, we evaluate the system performance under different values of . In another word, one time of successful detection is that the system detects times of smoking motions and the time intervals between any two continue detected motions are similar. Evaluation results are shown in Fig. 18. The system accuracy decreases a little with the increase of , while the false alarm rate first decreases sharply and then remains almost unchanged with the increase of . This is because the larger the , the more smoking motions need to continuously and correctly detect, and less probability of identifying other activities as smoking motions. We can find that when 3, the false alarm rate is low enough and with the increase of , false alarm rate almost no longer decreases. To keep a balance between accuracy and false alarm rate, we take ¼ 3 as the best choice.

### 5.2.8 Impact of Network Architecture and Dataset Size

The network architecture used in HearSmoking not only influences system performance, but also influences computation cost. Since the size of training dataset has impact on model training and further system accuracy, we design other three network structures to study the relationship between the training dataset size and the system performance. 3-CNN is a CNN consisting of one convolution layer, one fully connected layer and one softmax layer. 5-CNN is a CNN consisting of three convolution layers, one fully connected layer and one softmax layer. 3-DNN is a deep neural network with three fully connected layers. 4-CNN is the network in HearSmoking. The results are shown in Fig. 19. It is obvious that for each network, the system accuracy increases with the dataset size increases at beginning, and remains stable when the dataset size exceeds a specific value. DNN has the lowest accuracy indicating that it is not suitable for HearSmoking. 5-CNN has higher accuracy when dataset size exceeds 6,000 but its complex structure and high computational complexity make it inappropriate to be deployed on smartphones. Therefore, to achieve a high system performance while keep an acceptable computation cost, it is suitable to select 4-CNN and set dataset size as 5,000.

## 5.3 Energy Efficiency Evaluation

HearSmoking is deployed on smartphones, so power consumption is a problem which needs to be considered. Through our empirical study, we find that the power consumption of our system mainly includes two parts: one is the power consumed by training the CNN in offline phase, the other is the power consumed by detecting the driver's movements in online phase. The process of training CNN is not carried out on the driver's smartphone, so this part of the power consumption does not affect the driver's experience. We just need to focus on the power consumed by detecting driver's movements in online phase. It includes two parts: one is the power consumed by computation, the other is the power consumed by sensor operation. First, the most energy consumption from computation is during the CNN classification. However, existing works [31], [32] show that compared with other classification methods, neural networks can manage power consumption better. Second, as shown in [22], [33], compared with the power consumed by computation, the power consumed by sensor operation dominates the power consumption of a smartphone. Since speakers and microphones are the sensors in our system, we now study the impact of audio volume on power consumption and system performance.

We use 4 different models of smartphones and define the maximum volume of each smartphone as 100 percent. Then we conduct experiments within the volume range of 0 to 100 percent. Note that the volume of 0 percent represents all APPs and services on the smartphone are closed. Fig. 20 shows the power consumed by each smartphone running for one hour under different volume. We can see that as the volume increases, the power consumed by the smartphone also increases. In addition, we compare the accuracy of our system at different volume. We find that when the volume is greater than 50 percent, the system accuracy can exceed 90 percent. When the volume reaches 60 percent, our system achieves an average accuracy of 93.44 percent. Besides, the power consumption of 60 percent volume is 0.68 times of that of 100 percent volume. So we choose 60 percent volume as the default volume for all experiments.

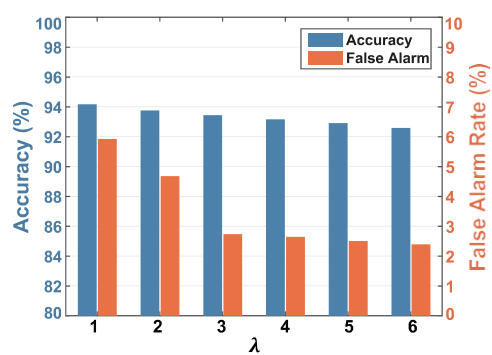

Fig. 18. Impact of period detection times .

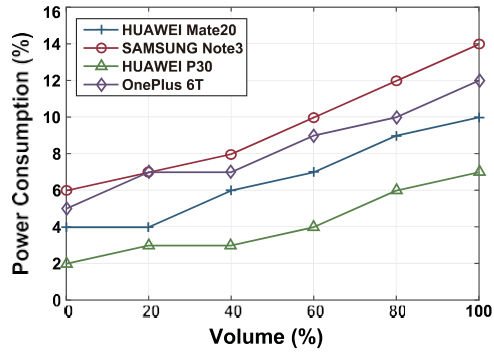

Fig. 20. Power consumption under different audio volume.

## 6  CONCLUSION

In this paper, we address how to detect cigarette smoking activity during driving to improve road safety. Through literature survey and experimental verification, we find some characteristic smoking patterns in driving environment. We propose a smoking detection system, named HearSmoking, which leverages acoustic sensors on smartphones to detect cigarette smoking events of drivers when they are driving. HearSmoking takes advantages of RCC and CNN to detect both hand movements and respirations of the driver. Methods of composite analysis and periodicity analysis are designed to improve system performance. We conduct extensive experiments in different driving environments. HearSmoking can detect smoking events with an average accuracy of 93.44 percent in real-time, which indicates that it works efficiently and reliably.

## ACKNOWLEDGMENTS

The work of Fan Li was supported in part by the National Natural Science Foundation of China (NSFC) under Grants 62072040, 61772077, and the Beijing Natural Science Foundation under Grant 4192051.

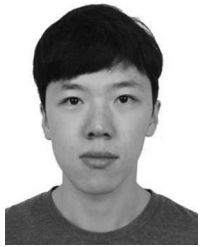

**Yadong Xie** received the BE degree in network engineering from Hebei University, China, in 2016. Currently, he is working toward the PhD degree from the School of Computer Science, Beijing Institute of Technology, Beijing, China. His research interests include Internet of Vehicles, mobile computing, and machine learning.

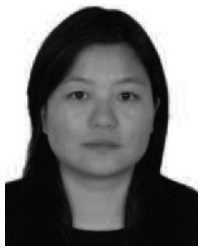

**Fan Li** (Member, IEEE) received the BEng and MEng degrees in communications and information system from the Huazhong University of Science and Technology, China, in 1998 and 2001, respectively, the MEng degree in electrical engineering from the University of Delaware, Newark, Delaware, in 2004, and the PhD degree in computer science from the University of North Carolina at Charlotte, Charlotte, North Carolina, in 2008. She is currently a professor at the School of Computer Science, Bei- jing Institute of Technology, China. Her current research interests include wireless networks, ad hoc and sensor networks, and mobile computing. Her papers won Best Paper Awards from IEEE MASS (2013), IEEE IPCCC (2013), ACM MobiHoc (2014), and Tsinghua

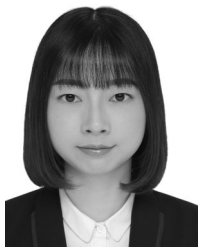

**Yue Wu** received the BE degree in Internet of Things from the Beijing Institute of Technology, China, in 2015. Currently, she is working toward the PhD degree from the School of Computer Science, Beijing Institute of Technology, Beijing, China. Her research interests include mobile crowd sensing, edge computing, and deep learning.

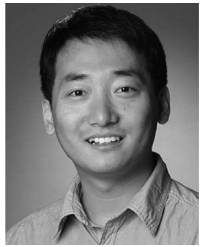

**Song Yang** (Member, IEEE) received the PhD degree from the Delft University of Technology, The Netherlands, in 2015. From August 2015 to July 2017, he worked as postdoc researcher for the EU FP7 Marie Curie Actions CleanSky Project in Gesell- schaft fu€r wissenschaftliche Datenverarbeitung mbH Go€ttingen (GWDG), Go€ttingen, Germany. He is currently an associate professor at the School of Computer Science, Beijing Institute of Technology, China. His research interests include data communi- cation networks, cloud/edge computing, and net- work

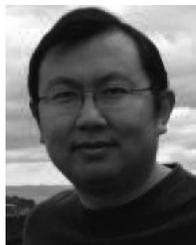

**Yu Wang** (Fellow, IEEE) received the BEng and MEng degrees from Tsinghua University, China, and the PhD degree from the Illinois Institute of Technology, Chicago, Illinois, all in computer science. He is currently a professor at the Department of Computer and Information Sciences, Temple University, Philadelphia, Pennsylvania. His research interests include wireless networks, smart sensing, and mobile computing. He has published more than 200 papers in peer reviewed journals and conferences, with four best paper awards. He has served as general chair, program chair, program committee member, etc. for many international conferences (such as IEEE IPCCC, ACM MobiHoc, IEEE INFOCOM, IEEE GLOBECOM, IEEE ICC). He has served as editorial board member of several international journals, including the *IEEE Transactions on Parallel and Distributed Systems*. He is a recipient of Ralph E. Powe Junior Faculty Enhancement Awards from Oak Ridge Associated Universities (2006), Outstanding Faculty Research Award from College of Computing and Informatics at UNC Charlotte (2008). He is also a senior member of the ACM and a member of the AAAS.

▸▸ For more information on this or any other computing topic, please visit our Digital Library at www.computer.org/csdl.